\documentclass[aps,pra,reprint,superscriptaddress,onecolumn]{revtex4-2}
\usepackage{amsmath}
\usepackage{amsfonts}
\usepackage{xcolor,graphicx}        
\usepackage{braket}
\usepackage{float}
\usepackage{tikz}
\usepackage{hyperref}
\usepackage{booktabs}  
\usepackage{multirow}  
\usepackage{amsmath}    
\usepackage{array}
\hypersetup{
	colorlinks = true,
	linkcolor = cyan,
	anchorcolor = cyan,
	citecolor = cyan,
	filecolor = cyan,
	urlcolor = cyan}
\begin{document}
\title{Paraxial beam propagation from {A}iry-type initial conditions via the {O}perator {M}ethod}

\author{I. Julían-Macías}
\affiliation{Instituto Nacional de Astrofísica Óptica y Electrónica (INAOE)\\ Luis Enrique Erro 1, Santa María Tonantzintla, Puebla, 72840, Mexico}

\author{M. A. Jácome-Silva}
\affiliation{Instituto Nacional de Astrofísica Óptica y Electrónica (INAOE)\\ Luis Enrique Erro 1, Santa María Tonantzintla, Puebla, 72840, Mexico}

\author{I. Ramos-Prieto}
\affiliation{Instituto Nacional de Astrofísica Óptica y Electrónica (INAOE)\\ Luis Enrique Erro 1, Santa María Tonantzintla, Puebla, 72840, Mexico}

\author{U. Ruiz-Corona}
\affiliation{Instituto Nacional de Astrofísica Óptica y Electrónica (INAOE)\\ Luis Enrique Erro 1, Santa María Tonantzintla, Puebla, 72840, Mexico}

\author{F. Soto-Eguibar}
\affiliation{Instituto Nacional de Astrofísica Óptica y Electrónica (INAOE)\\ Luis Enrique Erro 1, Santa María Tonantzintla, Puebla, 72840, Mexico}

\author{D. {Sánchez}-{de-la-Llave}}
\affiliation{Instituto Nacional de Astrofísica Óptica y Electrónica (INAOE)\\ Luis Enrique Erro 1, Santa María Tonantzintla, Puebla, 72840, Mexico}

\author{H. M. Moya-Cessa}
\affiliation{Instituto Nacional de Astrofísica Óptica y Electrónica (INAOE)\\ Luis Enrique Erro 1, Santa María Tonantzintla, Puebla, 72840, Mexico}

\begin{abstract}
We employ quantum mechanical operator techniques to solve the equations of $(1+1)D$ and $(2+1)D$ for paraxial waves with initial conditions defined by Airy-type functions. In the first part, we find the expressions of $(1+1)D$ optical beams, considering initial conditions such as Airy, Airy-truncated, and Airy-Gaussian functions. Subsequently, we extended the analysis to $(2+1)D$ optical beams with initial conditions generated by the product of two Airy, two Airy-truncated and two Airy-Gaussian functions, providing a comprehensive study of multidimensional Airy beam propagation. To validate our theoretical derivations, we present both theoretical and experimental intensity profiles, showing excellent agreement between the two, illustrating the physical characteristics of these beams. Although these solutions have previously been obtained via the diffraction integral and thoroughly studied, the primary goal here is to demonstrate that the optical fields can be derived using quantum mechanical operator methods. Finally, we remark that this alternative approach offers an elegant and powerful framework for analyzing paraxial wave propagation.
\end{abstract}

\maketitle

\section{Introduction}\label{Introduction}
One of the most elegant and pedagogically fruitful symmetries in undergraduate and graduate physics is the formal isomorphism between the paraxial wave equation in optics and the time-dependent Schrödinger equation for a free particle in quantum mechanics. Although students typically encounter these equations in completely separate courses, bridging them provides a profound conceptual advantage. Mathematical tools developed to describe the time evolution of quantum states can be directly translated to describe the spatial propagation of optical fields. In this paper, we exploit this analogy to introduce advanced undergraduate and graduate students to the power of quantum mechanical operator methods  by solving the propagation dynamics of paraxial Airy-type beams. 

The Airy wave packet serves as an ideal case study for this cross-disciplinary approach. Originally introduced by Berry and Balazs as a unique solution to the $(1+1)D$ free-particle Schrödinger equation \cite{BerryNonspreading}, the Airy wave packet exhibits the counterintuitive property of propagating without diffraction while experiencing constant transverse acceleration. This theoretical curiosity was later realized in the optical regime, deriving the corresponding solution for the $(1+1)D$ paraxial wave equation \cite{EspindolaParaxial, ZannottiThesis}. Furthermore, the energy flux \cite{SanzFlux, SanzExploring} and the Gouy phase \cite{PangTheyGouy} of a $(1+1)D$ Airy beam have been studied. Similarly, the behavior of the $(2+1)D$ Airy beam has also been documented \cite{BerryStable, ZannottiThesis}. Because the ideal Airy beam possesses infinite energy and is not square-integrable, Siviloglou and Christodoulides \cite{SiviloglouAccelerating, SiviloglouObservation} introduced finite energy variants, such as the Airy-truncated beam. These structured optical fields have sparked significant interest due to their  properties, such as ballistic motion \cite{SiviloglouBallistic, HuOptimal} self-healing\cite{BrokySelfhealing, ChuAnalytical, QianEvolution, AnayaAiry}, energy flux \cite{SanzFlux, SanzExploring, SztulThePoynting} and applications in optical trapping \cite{ZhengOptical, ZhaoOptical}. Another type of finite Airy beam is generated from an initial condition given by an Airy function multiplied by the Gaussian factor, termed an Airy-Gaussian beam \cite{BandresAiry-Gauss}. Additionally, the propagation of Airy, Airy-truncated, and Airy-Gaussian beams has also been studied in various media. For example, the propagation of a $(1+1)D$ Airy beam in dynamic linear index potentials \cite{EfremidisAiry}, the $(1+1)D$ Airy-truncated beam in GRIN linear and quadratic media \cite{AsenjoPropagation}, and the $(2+1)D$ Airy-truncated beam in GRIN quadratic media \cite{LiPropagation, ZhangInteraction}. Similarly, the propagation of the $(1+1)D$ Airy-Gaussian beam in GRIN linear media \cite{AsenjoPropagation, DengPropagation} and the $(2+1)D$ Airy-Gaussian beam in GRIN quadratic media \cite{LiuPropagation} has been investigated. Finally, we note that the preceding Airy-type beams have been generated using different methods; for example, solutions in free space are obtained using the Fourier transform method \cite{SiviloglouAccelerating, SiviloglouBallistic} and operator techniques used in quantum mechanics \cite{AnayaAiry}. On the other hand, for GRIN linear and quadratic media, the \textsc{ABCD} matrix method \cite{BandresAiry-Gauss} and operator techniques used in quantum mechanics \cite{AsenjoPropagation} have been applied. Traditionally, the propagation of these complex beams in free space is derived by using the Fresnel diffraction integral or Fourier transform methods. Although effective, evaluating these integrals for complex initial conditions such as the Airy-truncated or Airy-Gaussian profiles often obscures the underlying physics behind tedious calculus.

The primary pedagogical goal of this work is to demonstrate an alternative, purely algebraic framework based on quantum mechanical operator techniques. By treating the propagation distance as an evolution parameter , we show how optical fields can be elegantly derived using non-commuting operator algebra. We guide the reader through the application of the Hadamard lemma and the Baker-Campbell-Hausdorff formula —tools standardly taught in quantum mechanics— to factorize and evaluate the evolution of $(1+1)D$ and $(2+1)D$ optical beams. We remark that quantum mechanical operator techniques have been employed in order to introduce symmetric and asymmetric Cauchy-Riemann beams \cite{MoyaCauchy, KorneevAsymmetric} in free space, Cauchy-Riemann beams in GRIN linear and quadratic media \cite{RamosCauchy}, beams based on Bessel and Airy functions of complex variable \cite{MoyaParaxial}, demonstrating the feasibility of introducing new optical beams with specific characteristics. Similarly, it has been shown that with this method it is possible to obtain solutions that are already known, such as Gaussian apodized Whittaker integrals and Gaussian apodized Helmholtz fields \cite{SilvaOperator}. In the first part of this paper, we establish the mathematical framework for $(1+1)D$ paraxial waves, utilizing integral representations of Airy-ideal, Airy-truncated and Airy-Gaussian functions as initial conditions. We then extend this operator formalism to multidimensional $(2+1)D$ beams. To validate our theoretical derivations and provide a complete physical picture, we complement our analytical results with experimental intensity profiles obtained via a $4$-$f$ optical system, demonstrating to students the direct link between abstract operator algebra and observable laboratory phenomena. Ultimately, this approach not only simplifies the derivation of complex optical fields but also enriches the student's mathematical toolkit for analyzing wave propagation.

\section{Propagation of $(1+1)D$ paraxial beams using Airy-type functions as initial conditions}

In this section, we employ the operator formalism commonly used in quantum mechanics to analyze the propagation of optical beams governed by the $(1+1)$D paraxial wave equation in free space:
\begin{equation}
\frac{\partial^2 u(x,z)}{\partial x^2} + 2ik\frac{\partial u(x,z)}{\partial z} = 0.  
\label{pwe1D}
\end{equation}
Due to its formal analogy with the Schrödinger equation for a free particle, the model of operator techniques allows the use of unitary evolution operators to obtain the propagated initial condition. The solution of Eq.~(\ref{pwe1D}) can be expressed in the operator form as \cite{StolerOperator}
\begin{equation}
    u(x,\tau) = e^{\frac{i\tau}{2} \frac{\partial^2}{\partial x^2}} u(x,0),\label{propagator1D}
\end{equation}
where $\tau = z/k$ is a normalized propagation distance and $\exp\left(\frac{i\tau}{2} \frac{\partial^2}{\partial x^2}\right)$ is the spatial evolution operator. To illustrate this formalism, we consider the following initial condition
\begin{equation}
   u(x,0) = f(x) \int_{-\infty}^{\infty} e^{i\left(xP + \frac{P^3}{3}\right)}dP,\label{E(x,0)1D}
\end{equation}
with $f(x) = \{1,\, \exp(\alpha x),\, \exp(-g x^2)\}$, where $\alpha$ and $g$ are complex numbers representing the truncation factor and the transverse confinement factor, respectively; these profiles allow for the generation of structured optical fields with distinct propagation properties. We note that Eq.~(\ref{E(x,0)1D}) corresponds to the expression $ 2\pi f(x) \text{Ai}(x)$, since the integral is an integral representation of the Airy function $\text{Ai}(x)$ (times $2\pi$). In addition, for each expression of the function $f(x)$, we obtain the well-known $(1+1)D$ Airy-ideal, Airy-truncated and Airy-gauss beams, respectively, which have been extensively studied.

\subsection{Propagation of a paraxial beam with an Airy function as initial condition}\label{Airy-Ideal1D}

As mentioned in the Introduction, two methods were presented to obtain the $(1+1)D$ Airy beam \cite{EspindolaParaxial, ZannottiThesis}. The first method shows that $\exp[iS(x,z)]$ is a solution of the paraxial wave equation in free space if and only if $S(x,z)$ satisfies both the Hamilton–Jacobi and Laplace equations. The second method uses the well-known Fresnel transform. To find the expression of the $(1+1)D$ Airy beam in free space, using operator techniques, we consider Eq.~(\ref{E(x,0)1D}) as the initial condition with $f(x) = 1$, that is,
\begin{equation}
u_1(x,0) = \int_{-\infty}^{\infty} e^{i\left(xP + \frac{P^3}{3}\right)}dP,\label{E1(x,0)D_1}
\end{equation}
which corresponds to the Airy function given by $2\pi\mathrm{Ai}(x)$. Substituting the initial condition Eq.~(\ref{E1(x,0)D_1}) into Eq.~(\ref{propagator1D}), the propagated beam is given by  
\begin{equation}
u_1(x,\tau) = e^{\frac{i\tau}{2}\frac{\partial^2}{\partial x^2}} \int_{-\infty}^{\infty} e^{ixP} e^{\frac{iP^3}{3}}dP.\label{E1(x,tau)2D_1}
\end{equation}
We note that $\exp(ixP)$ is an eigenfunction of the operator $\frac{\partial^2}{\partial x^2}$ with eigenvalue $-P^2$, that is, $\frac{\partial^2}{\partial x^2} \exp(ixP) = -P^2 \exp(ixP)$; thus $\exp\left(\frac{i\tau}{2}\frac{\partial^2}{\partial x^2}\right) \exp(ixP) = \exp\left(-\frac{i\tau P^2}{2}\right) \exp(ixP)$. Using this last result, the expression for the propagated Airy beam is given by 
\begin{equation}
    u_1(x,\tau) = \int_{-\infty}^{\infty} e^{i \left( xP - \frac{ \tau P^2}{2} + \frac{P^3}{3} \right)}dP,\label{E1(x,tau)3D_1}
\end{equation}
which can be expressed in terms of the Airy function as
\begin{equation}
    u_1(x,\tau) = 2\pi \mathrm{Ai}\left(x - \frac{\tau^2}{4} \right) e^{i \left( \frac{\tau x}{2} - \frac{ \tau^3}{12} \right)}.\label{Airyidealbeam}
\end{equation}
We mention that Eq.~(\ref{Airyidealbeam}) is in agreement with the findings reported in \cite{EspindolaParaxial, ZannottiThesis}, with only slight variations in the definitions and notation of the variables. Finally, Appendix \ref{app:alternative} shows how to obtain Eq.~(\ref{E1(x,tau)3D_1}) using the Hadamard lemma and the Baker-Campbell-Hausdorff formula.

\subsection{Propagation of a paraxial beam with an Airy-truncated function as initial condition}\label{Airy-Truncated1D}

The finite-energy Airy beam introduced in \cite{SiviloglouAccelerating, SiviloglouObservation}, known as the Airy-truncated beam, is obtained by the Fresnel transform. To derive its expression, we take as an initial condition Eq.~(\ref{E(x,0)1D}) with $f(x) = \exp(\alpha x)$, which leads to
\begin{equation}
u_2(x,0) = e^{\alpha x} \int_{-\infty}^{\infty} e^{i\left(xP + \frac{P^3}{3}\right)}dP,\label{E2(x,0)D_1}
\end{equation}
recognized as $\exp(\alpha x) \mathrm{Ai}(x)$.
Substituting Eq.~(\ref{E2(x,0)D_1}) into Eq.~(\ref{propagator1D}), we obtain the propagated beam in the form of
\begin{equation}
u_2(x,\tau) = e^{\frac{i\tau}{2}\frac{\partial^2}{\partial x^2}} e^{\alpha x} \int_{-\infty}^{\infty} e^{i\left(xP + \frac{P^3}{3}\right)}dP.\label{E2(x,tau)2D_1}
\end{equation}
By inserting the identity operator $\hat{I} = \exp\left(-\frac{i\tau}{2}\frac{\partial^2}{\partial x^2}\right) \exp\left( \frac{i\tau}{2}\frac{\partial^2}{\partial x^2}\right)$ after the factor $\exp(\alpha x)$, we rewrite Eq. (\ref{E2(x,tau)2D_1}) as
\begin{equation}
u_2(x,\tau) = e^{\frac{i\tau}{2}\frac{\partial^2}{\partial x^2} } e^{\alpha x} e^{-\frac{i\tau}{2}\frac{\partial^2}{\partial x^2}} e^{\frac{i\tau}{2}\frac{\partial^2}{\partial x^2}} \int_{-\infty}^{\infty} e^{i\left(xP + \frac{P^3}{3}\right)}dP.\label{E2(x,tau)3D_1}
\end{equation}
This expression can be interpreted as the product of $\exp\left(\frac{i\tau}{2}\frac{\partial^2}{\partial x^2}\right) \exp(\alpha x) \exp\left(-\frac{i\tau}{2}\frac{\partial^2}{\partial x^2}\right)$ and $\exp\left(\frac{i\tau}{2}\frac{\partial^2}{\partial x^2}\right)$ acting on the integral. To simplify the first factor, we apply the Hadamard lemma and the Baker-Campbell-Hausdorff formula \cite{Rossmann2002, Hall2013}, which yields the following
\begin{eqnarray}
e^{\frac{i\tau}{2}\frac{\partial^2}{\partial x^2}} e^{\alpha x} e^{-\frac{i\tau}{2}\frac{\partial^2}{\partial x^2}} & = & e^{\alpha\big( x + i\tau \frac{\partial}{\partial x}\big)} = e^{\frac{i \alpha^2 \tau}{2}} e^{\alpha x} e^{i\alpha \tau \frac{\partial}{\partial x}},
\end{eqnarray}
while the second factor corresponds exactly to Eq.~(\ref{E1(x,tau)2D_1}). Thus, Eq.~(\ref{E2(x,tau)3D_1}) becomes
\begin{equation}
u_2(x,\tau) = e^{\frac{i \alpha^2 \tau}{2}} e^{\alpha x} \int_{-\infty}^{\infty} e^{i\alpha \tau \frac{\partial}{\partial x}} e^{ixP} e^{i\left( - \frac{\tau P^2}{2} + \frac{P^3}{3} \right)}dP.\label{E2(x,tau)4D_1}
\end{equation}
To continue, we use the fact that $\exp\left(i\alpha \tau \frac{\partial}{\partial x}\right) \exp(ixP) = \exp(-\alpha \tau P) \exp(ixP)$. Then, regrouping the terms yields the propagated paraxial beam for the initial condition Eq.~(\ref{E2(x,0)D_1}):
\begin{equation}
u_2(x,\tau) = e^{\frac{i \alpha^2 \tau}{2}} e^{\alpha x} \int_{-\infty}^{\infty} e^{i\left[\big( x + i\alpha \tau \big)P - \frac{\tau P^2}{2} + \frac{P^3}{3}\right]}dP.\label{E2(x,tau)5D_1}
\end{equation}
In terms of the Airy function, this becomes
\begin{equation}
u_2(x,\tau) = 2\pi e^{\alpha x} e^{-\frac{\alpha \tau^2}{2}} \mathrm{Ai}\left(x + i \alpha \tau - \frac{\tau^2}{4} \right) e^{i \left( \frac{\tau x}{2} - \frac{\tau^3}{12} \right)} {e^{\frac{i \alpha^2 \tau}{2}}.}\label{Airytruncatedbeam}
\end{equation}
From Eq.~(\ref{Airytruncatedbeam}), it is clear that when $\alpha = 0$ the result reduces to Eq.~(\ref{Airyidealbeam}). For real $\alpha$, the expression coincides with that obtained in \cite{SiviloglouAccelerating}. Furthermore, for complex $\alpha$, it resembles the result reported in \cite{SiviloglouBallistic}, except for a difference in the scaling of the variable.

\subsection{Propagation of a paraxial beam with an Airy-Gaussian function as initial condition}\label{Airy-Gaussian1D}

Finally, another finite-energy Airy beam is the so-called Airy-Gauss beam, which was first derived in \cite{BandresAiry-Gauss} using the matrix method $\mathrm{ABCD}$. In order to derive the expression of the propagated beam, we take as an initial condition Eq.~(\ref{E(x,0)1D}) with $f(x) = \exp(-gx^2)$, namely,
\begin{equation}
   u_3(x,0) = e^{-g x^2} \int_{-\infty}^{\infty} e^{i\left(xP + \frac{P^3}{3}\right)}dP,\label{E3(x,0)D_1} 
\end{equation}
which can be expressed as $\exp(-gx^2) \mathrm{Ai}(x)$. Substituting Eq.~(\ref{E3(x,0)D_1}) into Eq.~(\ref{propagator1D}), we obtain 
\begin{equation}
   u_3(x,\tau) = e^{\frac{i\tau}{2}\frac{\partial^2}{\partial x^2}} e^{-g x^2} \int_{-\infty}^{\infty} e^{i\left(xP + \frac{P^3}{3}\right)}dP.\label{E3(x,tau)D_1}
\end{equation}
Introducing the identity operator $\hat{I} = \exp\left(-\frac{i\tau}{2}\frac{\partial^2}{\partial x^2}\right) \exp\left( \frac{i\tau}{2}\frac{\partial^2}{\partial x^2}\right)$ after the factor $\exp(-gx^2)$ yields
\begin{equation}
   u_3(x,\tau) = e^{\frac{i\tau}{2}\frac{\partial^2}{\partial x^2}} e^{-g x^2} e^{-\frac{i\tau}{2}\frac{\partial^2}{\partial x^2}} e^{\frac{i\tau}{2}\frac{\partial^2}{\partial x^2}} \int_{-\infty}^{\infty} e^{i\left(xP + \frac{P^3}{3}\right)}dP.\label{E3(x,tau)2D_1}
\end{equation}
This expression corresponds to the product of $\exp\left(\frac{i\tau}{2}\frac{\partial^2}{\partial x^2}\right) \exp(-gx^2) \exp\left(-\frac{i\tau}{2}\frac{\partial^2}{\partial x^2}\right)$ and $\exp\left(\frac{i\tau}{2}\frac{\partial^2}{\partial x^2}\right)$ applied to the integral. To rewrite the first factor, we apply the Hadamard lemma, the Baker-Campbell-Hausdorff formula~\cite{Rossmann2002, Hall2013}, and the Wei-Norman theorem~\cite{Wei1963, Wei1964} (see Appendix \ref{app:disentangling}), and we obtain
\begin{equation}
    e^{\frac{i\tau}{2}\frac{\partial^2}{\partial x^2}} e^{-g x^2} e^{-\frac{i\tau}{2}\frac{\partial^2}{\partial x^2}} = e^{h_1x^2} e^{h_2\left(x\frac{\partial}{\partial x} + \frac{\partial}{\partial x}x \right)} e^{h_3\frac{\partial^2}{\partial x^2}}.\label{DisentanglingIdentity}
\end{equation}
Substituting the last expression into Eq.~(\ref{E3(x,tau)2D_1}) gives
\begin{equation}
   u_3(x,\tau) = e^{h_1 x^2} \int_{-\infty}^{\infty} e^{h_2\left(x\frac{\partial}{\partial x} + \frac{\partial}{\partial x}x \right)} e^{ixP} e^{-h_2\left(x\frac{\partial}{\partial x} + \frac{\partial}{\partial x}x \right)} e^{h_2\left(x\frac{\partial}{\partial x} + \frac{\partial}{\partial x}x \right)} e^{i \left[- \left( \frac{\tau}{2} - i h_3 \right)P^2 + \frac{P^3}{3} \right]}dP,\label{E3(x,tau)4D_1}
\end{equation}
from the Taylor expansion of the exponential operator $\exp\big(h_3\frac{\partial^2}{\partial x^2}\big)$ one obtains that $\exp\big(h_3\frac{\partial^2}{\partial x^2}\big) \exp(ixP) = \exp(-h_3 P^2) \exp(ixP)$ and have placed the identity operator $\hat{I} = \exp\big[-h_2\big(x\frac{\partial}{\partial x} + \frac{\partial}{\partial x}x \big)\big] \exp\big[h_2\big(x\frac{\partial}{\partial x} + \frac{\partial}{\partial x}x \big)\big]$ after the factor $\exp(ixP)$. In conclusion, because $\exp\big[h_2\big(x\frac{\partial}{\partial x} + \frac{\partial}{\partial x}x \big)\big]\exp\big\{i\big[-\big( \frac{\tau}{2} - ih_3 \big) P^2 + \frac{P^3}{3} \big]\big\} = \exp(h_2) \exp\big\{i\big[-\big( \frac{\tau}{2} - ih_3 \big) P^2 + \frac{P^3}{3} \big]\big\}$, and given that $\exp\big[h_2\big(x\frac{\partial}{\partial x} + \frac{\partial}{\partial x}x \big)\big] \exp(ixP) \exp\big[-h_2\big(x\frac{\partial}{\partial x} + \frac{\partial}{\partial x}x \big)\big] = \exp\big[ix \exp(2h_2) P\big]$, by rearranging the terms and taking into account the functions $h_s$ given by Eq.~(\ref{functionshs}), we derive the ultimate form of the paraxial beam
\begin{equation}
   u_3(x,\tau) = \frac{e^{-\frac{gx^2}{\omega(\tau)}}}{\sqrt{\omega(\tau)}} \int_{-\infty}^{\infty} e^{i\left[ \frac{x}{\omega(\tau)}P - \left( \frac{\tau}{2}- \frac{ig\tau^2}{\omega(\tau)} \right) P^2 + \frac{P^3}{3} \right]}dP,\label{E3(x,tau)7D_1}
\end{equation}
which, using the integral representation of the Airy function, can be written as
\begin{equation}
    u_3(x,\tau) = 2\pi \frac{e^{\frac{-gx^2}{\omega(\tau)}}}{\sqrt{\omega(\tau)}}  \mathrm{Ai}\left[\frac{x}{\omega(\tau)} - \left(\frac{\tau}{2} - \frac{ig\tau^2}{\omega(\tau)} \right)^2 \right] e^{i\left[ \frac{x}{\omega(\tau)} \left( \frac{\tau}{2} - \frac{ig\tau^2}{\omega(\tau)} \right) - \frac{2}{3}\left( \frac{\tau}{2}- \frac{ig\tau^2}{\omega(\tau)} \right)^3 \right]}.\label{Airygaussianbeam}
\end{equation}
Note that Eq.~(\ref{Airygaussianbeam}) reduces to  Eq.~(\ref{Airyidealbeam}) when $g = 0$. Furthermore, in this expression, the Gaussian factor $\exp\big[-\frac{g x^{2}}{\omega(\tau)}\big]$ appears, which clearly identifies the solution as an Airy-Gaussian beam. The denominator $1/\sqrt{\omega(\tau)}$ is typical of Gaussian solutions under propagation, ensuring both normalization and transverse spreading, where the parameter $g$ acts as a confinement factor, making the beam more localized and physically realizable than in the exponential case Eq.~(\ref{Airytruncatedbeam}). The structure inside the Airy argument still exhibits the characteristic self-accelerating parabolic trajectory but is now modulated by the Gaussian confinement.

\section{Propagation of $(2+1)D$ paraxial beams with Airy-type functions as initial conditions}

In this section, we employ the operator formalism commonly used in quantum mechanics to analyze the propagation of optical beams governed by the $(2+1)$D paraxial wave equation in free space:
\begin{equation}
   \frac{\partial^2 U(x,y,z)}{\partial x^2} + \frac{\partial^2 U(x,y,z)}{\partial y^2} +2ik\frac{\partial U(x,y,z)}{\partial z}  = 0.\label{pwe2D}
\end{equation}
Due to its formal analogy with the Schrödinger equation, this model allows the use of unitary evolution operators to describe the field dynamics. The general solution to Eq.~(\ref{pwe2D}) can be expressed in operator form as \cite{StolerOperator}
\begin{equation}
    U(x,y,\tau) = e^{\frac{i\tau}{2} \left( \frac{\partial^2}{\partial x^2} +\frac{\partial^2}{\partial y^2} \right) }U(x,y,0),\label{propagator2D}
\end{equation}
where $\tau = z/k$. Therefore, the propagation of the optical beams is obtained when the evolution operator $\exp\big[\frac{i\tau}{2}\big( \frac{\partial^2}{\partial x^2} + \frac{\partial^2}{\partial y^2}\big)\big]$ is applied to the initial condition. To exemplify this method, we consider the following expression as an initial condition
\begin{equation}
   U(x,y,0) = f(x,y) \int_{-\infty}^{\infty} e^{i\left(xP + P^3/3\right)}dP \int_{-\infty}^{\infty} e^{i\left(yQ + Q^3/3\right)}dQ,\label{E(x,y,0)2D}
\end{equation}
with $f(x,y) = \{1,\, \allowbreak \exp[\alpha (x + y)],\, \allowbreak \exp[-g (x^2 + y^2)] \}$. Observe that this initial condition corresponds to $4\pi^2 f(x) \text{Ai}(x) \allowbreak f(y) \text{Ai}(y)$. In addition, for each expression of the function $f(x,y)$, we obtain the well-known $(2+1)D$ Airy-ideal, Airy-truncated and Airy-Gaussian beams, respectively, which are widely studied.

\subsection{Propagation of paraxial beams with an Airy function as initial condition}\label{Airy-Ideal2D}

The $(2+1)D$ Airy beam was obtained by using the Fresnel transform \cite{BerryStable, ZannottiThesis}. To obtain the expression of this optical beam, we consider the first case as the initial condition given by Eq.~(\ref{E(x,y,0)2D}), considering $f(x,y) = 1$, which can be expressed as  
\begin{equation}
    U_1(x,y,0) = \int_{-\infty}^{\infty} e^{i\left(xP + P^3/3\right)}dP \int_{-\infty}^{\infty} e^{i\left(yQ + Q^3/3\right)}dQ.\label{E1(x,y,0)D_2}
\end{equation}
Substituting Eq.\,(\ref{E1(x,y,0)D_2}) into Eq.~(\ref{propagator2D}), we obtain 
\begin{equation}
   U_1(x,y,\tau) = e^{\frac{i\tau}{2} \left( \frac{\partial^2}{\partial x^2} + \frac{\partial^2}{\partial y^2} \right)} \int_{-\infty}^{\infty} e^{i\left(xP + \frac{P^3}{3}\right)}dP \int_{-\infty}^{\infty} e^{i\left(yQ + \frac{P^3}{3}\right)}dQ.\label{E1(x,y,0)D_1}
\end{equation}
To apply the operator $\exp\left[\frac{i\tau}{2} \left( \frac{\partial^2}{\partial x^2} + \frac{\partial^2}{\partial y^2} \right)\right]$ on a function given by $F(x)F(y)$, it is necessary to use the Baker–Campbell\\–Hausdorff formula on the spatial evolution operator; that is, $\exp\left[\frac{i\tau}{2} \left( \frac{\partial^2}{\partial x^2} + \frac{\partial^2}{\partial y^2} \right)\right] = \exp\left(\frac{i\tau}{2} \frac{\partial^2}{\partial x^2}\right) \exp\left(\frac{i\tau}{2}\frac{\partial^2}{\partial y^2}\right)$. With this result,  Eq.~(\ref{E1(x,y,0)D_1}) takes the form
\begin{equation}
     U_1(x,y,\tau) = e^{\frac{i\tau}{2} \frac{\partial^2}{\partial x^2}} e^{\frac{i\tau}{2} \frac{\partial^2}{\partial y^2}} \int_{-\infty}^{\infty} e^{i\left(xP + \frac{P^3}{3}\right)}dP \quad e^{-\frac{i\tau}{2} \frac{\partial^2}{\partial y^2}} e^{\frac{i\tau}{2} \frac{\partial^2}{\partial y^2}} \int_{-\infty}^{\infty} e^{i\left(yQ + \frac{P^3}{3}\right)}dQ,\label{E1(x,y,0)2D_1}
\end{equation}
where we placed the identity operator $\hat{I} = \exp\left(-\frac{i\tau}{2}\frac{\partial^2}{\partial y^2}\right) \exp\left(\frac{i\tau}{2}\frac{\partial^2}{\partial y^2}\right)$ after the integral with respect to $P$. It is important to note that by applying the Hadamard lemma \cite{Rossmann2002, Hall2013}, we obtain $\exp\left(\frac{i\tau}{2}\frac{\partial^2}{\partial y^2}\right) F(x) \exp\left(\frac{i\tau}{2}\frac{\partial^2}{\partial y^2}\right) = F(x)$, which yields 
\begin{equation}
     U_1(x,y,\tau) = e^{\frac{i\tau}{2} \frac{\partial^2}{\partial x^2}} \int_{-\infty}^{\infty} e^{ixP} e^{i\frac{P^3}{3}}dP \quad  e^{\frac{i\tau}{2} \frac{\partial^2}{\partial y^2}} \int_{-\infty}^{\infty} e^{iyQ} e^{i\frac{Q^3}{3}}dQ.\label{E1(x,y,0)3D_1}
\end{equation}
We can see that the final expression can be written as the product of two terms. Each term has the same structure as Eq.~(\ref{E1(x,tau)2D_1}), which is expressed in the form of Eq.~(\ref{Airyidealbeam}). Using this observation, we finally obtain the expression for the Airy beam, given by 
\begin{eqnarray}
     U_1(x,y,\tau) & = & 4\pi^2 \mathrm{Ai}\left(x - \frac{\tau^2}{4} \right) \mathrm{Ai}\left(y - \frac{\tau^2}{4} \right) e^{i \left( \frac{\tau(x + y)}{2} - \frac{ \tau^3}{6} \right)}.\label{Airyidealbeam2D}
\end{eqnarray}
 
This result is the same as in \cite{BerryStable,ZannottiThesis}, except for small changes in notation and variables. To compare the algebraic operator formalism with the physical realization, we implemented an optical setup designed to synthesize and observe the propagation of Airy-type beams. This experiment serves as a practical demonstration of how a complex initial field distribution—represented mathematically as an initial state $u(x,y,0)$—evolves spatially according to the paraxial wave equation. The main device of the experimental setup is a reflective Spatial Light Modulator (SLM), which acts as the "programmable laboratory" for our initial conditions. The SLM allows us to display a specific phase modulation onto an incident plane wave. In our case, we used an expanded and collimated He-Ne laser ($\lambda = 632.8$ nm) to ensure a uniform phase front.  
For the experimental generation of Airy-type beams, a synthetic phase hologram (SPH) \cite{Arrizon:07} was encoded onto the SLM. As discussed in the operator derivation, the "truncation" or "Gaussian" envelopes are not just mathematical conveniences to ensure finite energy; in the laboratory, they represent the finite aperture of our SLM and the Gaussian profile of the laser source; this connection allows students to see the physical origin of the parameters $\alpha$ and $w_0$ (where $w_0=1/g)$ used in our equations. The propagation of the beam was analyzed using a 4-$f$ optical system, a standard tool in Fourier optics that is exceptionally instructive for students. A linearly polarized He-Ne laser beam illuminates a spatial light modulator (SLM, Holoeye), where the SPH is displayed. The reflected light propagates through the 4-$f$ system, which consists of two lenses with equal focal lengths ($f_1 = f_2 = 40 \text{ cm}$) and a spatial filter. The first lens performs the Fourier transform of the SPH in its focal plane, where the filter spatially transmits the desired field. Subsequently, the second lens performs the inverse Fourier transform of this filtered field, whose final intensity distribution is captured by a CCD camera.\\

\begin{figure}[!ht]
  \centering
  \includegraphics[width=\linewidth]{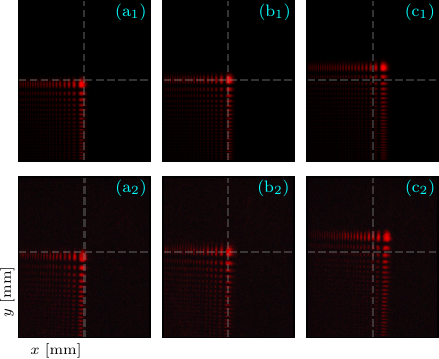}
  \caption{Intensity distributions for the Airy ideal beam at three transverse planes: ($\mathrm{a}_1$) at $z = 0.0 \, \mathrm{m}$, ($\mathrm{b}_1$) at $z = 0.2 \, \mathrm{m}$ and ($\mathrm{c}_1$) at $z = 0.4 \, \mathrm{m}$. The corresponding experimental distributions are shown in ($\mathrm{a}_2$)-($\mathrm{c}_2$), the experimental parameters are $\lambda = 632.8 \, \mathrm{nm}$ and $T = 1 \times 10^{-4} \, \mathrm{m}$ all within a viewing window of $4 \, \mathrm{mm}$. The excellent agreement validates the algebraic operator approach as a powerful alternative to the standard diffraction integral. Note the preservation of the beam's structure despite the transverse displacement, illustrating the non-diffracting nature of the Airy wave packet in a laboratory setting.}
  \label{figplotAiryi2D}
\end{figure}

This setup is the physical manifestation of the evolution operator $\hat{U}(\tau) = \exp(-\frac{i\tau}{2} \nabla^2)$ (with $\tau = z/k$ and $k = 2\pi/\lambda$ being the wavenumber). As we move the camera, we effectively ``apply" the operator to the initial state and record the resulting intensity distribution $I(x,y,z) = |U(x,y,\tau)|^2$. In Fig. \ref{figplotAiryi2D}, we present the analytical ($\mathrm{a}_1$)-($\mathrm{c}_1$) and experimental ($\mathrm{a}_2$)-($\mathrm{c}_2$) intensity distribution associated with Eq.~(\ref{Airyidealbeam2D}) in different planes given by different values of $z$, Furthermore, to generate the optical fields described in Eq.~(\ref{Airyidealbeam2D}), the parameter $T = 1 \times 10^{-4}\text{ m}$ is introduced; this value, also known as the characteristic Airy transverse scale $(x_0)$ \cite{SiviloglouAccelerating, SiviloglouObservation}, rescales the spatial variables $x$ and $y$. Similarly, the propagation coordinate $z$ is rescaled by $1/T^{2}$  defining the normalized distance $\tau_n = z/(k x_0^2)$ \cite{SiviloglouAccelerating, SiviloglouObservation}. The inclusion of the characteristic length $T$ is fundamental for generating Airy optical fields, as it makes the beam's description independent of the wavelength and aperture size; if this parameter were not used, the observation window would be completely illuminated. Additionally, given the large magnitude of $k$, the characteristic diffraction distance $kT^2$ proves to be very useful, since if the field were to propagate over distances $z$ much smaller than this magnitude $k$, the variation in the intensity profile would be imperceptible. Visually, the beam would appear static, as if it remained in its initial condition ($z=0$). Therefore, the use of $T$ (and consequently $T^2$) acts as an indispensable scaling factor that makes the evolution of Airy beams observable, allowing us to adequately analyze the propagation of the analytical solution for any value of $z$. As shown in Fig.~\ref{figplotAiryi2D}, there is a remarkable agreement between the simulated and experimental intensity distributions. Furthermore, a parabolic shift is observed in the $(x,y)$ plane as the beam propagates along the $z$-axis. 

\subsection{Propagation of paraxial beam with an Airy-truncated function as initial condition}\label{Airy-Truncated2D}

\begin{figure}[!ht]
  \centering
  \includegraphics[width=\linewidth]{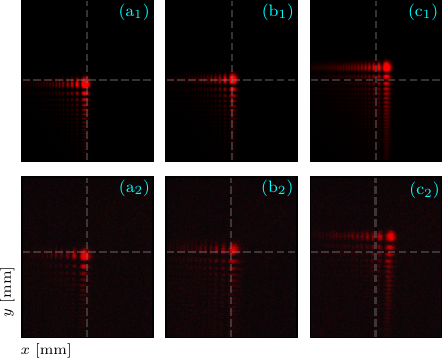}
  \caption{Intensity distributions for the Airy-truncated beam at three transverse planes: ($\mathrm{a}_1$) at $z = 0.0 \, \mathrm{m}$, ($\mathrm{b}_1$) at $z = 0.2  \, \mathrm{m}$ and ($\mathrm{c}_1$) at $z = 0.4 \, \mathrm{m}$. The corresponding experimental distributions are shown in ($\mathrm{a}_2$)-($\mathrm{c}_2$), the experimental parameters are $T = 1 \times 10^{-4} \, \mathrm{m}$, $\alpha=0.05$ and $\lambda = 632.8 \, \mathrm{nm}$, all within a viewing window of $4 \, \mathrm{mm}$. Unlike the ideal case, the introduction of the decay factor ensures the beam is square-integrable and physically realizable, resulting in a spatially bounded intensity distribution.}
  \label{figplotAiryt2D}
\end{figure}

As a second case, we examine the initial condition given by Eq.~(\ref{E(x,y,0)2D}) with $f(x,y) = \exp\left[\alpha(x+y)\right]$, which can be represented by
\begin{equation}
    U_2(x,y,0) = e^{\alpha x}\int_{-\infty}^{\infty} e^{i\left(xP + P^3/3\right)}dP\quad e^{\alpha y} \int_{-\infty}^{\infty} e^{i\left(yQ + Q^3/3\right)}dQ.\label{E2(x,y,0)D_2}
\end{equation}
Substituting Eq.~(\ref{E2(x,y,0)D_2}) into Eq.~(\ref{propagator2D}), we find that
\begin{equation}
   U_2(x,y,\tau) = e^{\frac{i\tau}{2} \left( \frac{\partial^2}{\partial x^2} + \frac{\partial^2}{\partial y^2} \right)} e^{\alpha x} \int_{-\infty}^{\infty} e^{i\left(xP + \frac{P^3}{3}\right)}dP\quad e^{\alpha y}\int_{-\infty}^{\infty} e^{i\left(yQ + \frac{P^3}{3}\right)}dQ.\label{E2(x,y,t)D_2}
\end{equation}
Applying the Baker-Campbell-Hausdorff formula \cite{Rossmann2002, Hall2013} to the evolution operator, Eq.~(\ref{E2(x,y,t)D_2}) can be expressed as
\begin{equation}
     U_2(x,y,\tau) = e^{\frac{i\tau}{2} \frac{\partial^2}{\partial x^2}} e^{\frac{i\tau}{2} \frac{\partial^2}{\partial y^2}} e^{\alpha x} \int_{-\infty}^{\infty} e^{i\left(xP + \frac{P^3}{3}\right)}dP \quad e^{-\frac{i\tau}{2} \frac{\partial^2}{\partial y^2}} e^{\frac{i\tau}{2} \frac{\partial^2}{\partial y^2}} e^{\alpha y} \int_{-\infty}^{\infty} e^{i\left(yQ + \frac{P^3}{3}\right)}dQ,
\end{equation}
where we located an identity operator $\hat{I} = \exp\left[-\frac{i\tau}{2} \frac{\partial^2}{\partial y^2}\right] \exp\left[\frac{i\tau}{2} \frac{\partial^2}{\partial y^2}\right]$ after the integral with respect to $P$. It is important to note that, using the Hadamard lemma, we obtain $\exp\left(\frac{i\tau}{2} \frac{\partial^2}{\partial y^2}\right) F(x) \exp\left(-\frac{i\tau}{2} \frac{\partial^2}{\partial y^2}\right) = F(x)$, so 
\begin{equation}
     U_2(x,y,\tau) = e^{\frac{i\tau}{2} \frac{\partial^2}{\partial x^2}} e^{\alpha x} \int_{-\infty}^{\infty} e^{i\left(xP + \frac{P^3}{3}\right)}dP \quad e^{\frac{i\tau}{2} \frac{\partial^2}{\partial y^2}} e^{\alpha y} \int_{-\infty}^{\infty} e^{i\left(yQ + \frac{P^3}{3}\right)}dQ.
\end{equation}
It is evident that the final expression can be represented as the product of two factors. Each factor shares the same form as Eq.~(\ref{E2(x,tau)2D_1}), and thus can be rewritten in the form of Eq.~(\ref{Airytruncatedbeam}). Based on this reasoning, we ultimately arrive at the expression for the Airy-truncated beam, given by
\begin{equation}
     U_2(x,y,\tau) = 4\pi^2 e^{\alpha (x + y)} e^{-\alpha \tau^2} \mathrm{Ai}\left(x + i \alpha \tau - \frac{\tau^2}{4} \right) \mathrm{Ai}\left(y + i \alpha \tau - \frac{\tau^2}{4} \right) e^{i\left( \frac{\tau (x + y)}{2} - \frac{\tau^3}{6} \right)} e^{i \alpha^2 \tau}.\label{Airytruncatedbeam2D}
\end{equation}
Note that when $\alpha = 0$, this expression reduces to Eq.~(\ref{Airyidealbeam2D}). Furthermore, we note that Siviloglou \cite{SiviloglouThesis} was the first to obtain the expression of the $(2+1)D$ Airy-truncated beam. The difference between his expression and Eq.~(\ref{Airytruncatedbeam2D}) lies in the rescaling of the variables. To generate this optical field, we employed the same experimental setup described in Section \ref{Airy-Ideal2D}, updating the synthetic phase hologram (SPH) displayed on the spatial light modulator (SLM) to correspond to this specific field. In Fig.~\ref{figplotAiryt2D}, we present the analytical and experimental intensity distributions associated with Eq.~(\ref{Airytruncatedbeam2D}) in various planes defined by different values of $\tau$. The intensity distributions reveal a clear contrast between ideal Airy beams and their truncated counterparts. Although the intensity of the ideal field possesses an infinite extent that does not vanish, the truncated version exhibits a spatially bounded distribution whose intensity decays to zero. Nevertheless, both solutions share the same propagation dynamics, manifesting a parabolic shift in the $(x,y)$ plane.

\subsection{Propagation of paraxial beam with an Airy-Gaussian function as initial condition}\label{Airy-Gaussian2D}
Finally, in the third case, we take the initial condition Eq.~(\ref{E(x,y,0)2D}) with $f(x,y) = \exp\left[-g(x^2+y^2)\right]$, and then the initial field can be written as  
\begin{equation}
    U_3(x,y,0) = e^{-gx^2}\int_{-\infty}^{\infty} e^{i\left(xP + P^3/3\right)}dP\quad e^{-gy^2} \int_{-\infty}^{\infty} e^{i\left(yQ + Q^3/3\right)}dQ.  \label{E3(x,y,0)D_2}
\end{equation}
Inserting Eq.~(\ref{E3(x,y,0)D_2}) into Eq.~(\ref{propagator2D}), we find
\begin{equation} 
   U_3(x,y,\tau) = e^{\frac{i\tau}{2} \left( \frac{\partial^2}{\partial x^2} + \frac{\partial^2}{\partial y^2} \right)} e^{-gx^2} \int_{-\infty}^{\infty} e^{i\left(xP + \frac{P^3}{3}\right)}dP\quad e^{-gy^2}\int_{-\infty}^{\infty} e^{i\left(yQ + \frac{P^3}{3}\right)}dQ.\label{0440}
\end{equation}
Applying once again the Baker-Campbell-Hausdorff formula \cite{Rossmann2002, Hall2013} to the evolution operator, Eq. (\ref{0440}) can be expressed as
\begin{equation}
     U_3(x,y,\tau) = e^{\frac{i\tau}{2} \frac{\partial^2}{\partial x^2}} e^{\frac{i\tau}{2} \frac{\partial^2}{\partial y^2}} e^{-gx^2} \int_{-\infty}^{\infty} e^{i\left(xP + \frac{P^3}{3}\right)}dP \quad e^{-\frac{i\tau}{2} \frac{\partial^2}{\partial y^2}} e^{\frac{i\tau}{2} \frac{\partial^2}{\partial y^2}} e^{-gy^2} \int_{-\infty}^{\infty} e^{i\left(yQ + \frac{P^3}{3}\right)}dQ,\label{E3(x,y,0)2D_2}
\end{equation}
where we located an identity operator $\hat{I} = \exp\left(-\frac{i\tau}{2} \frac{\partial^2}{\partial y^2}\right) \exp\left(\frac{i\tau}{2} \frac{\partial^2}{\partial y^2}\right)$ after the integral with respect to $P$. Using the Hadamard lemma, we obtain $\exp\left(\frac{i\tau}{2} \frac{\partial^2}{\partial y^2}\right) F(x) \exp\left(-\frac{i\tau}{2} \frac{\partial^2}{\partial y^2}\right) = F(x)$, so 
\begin{equation}
     U_3(x,y,\tau) = e^{\frac{i\tau}{2} \frac{\partial^2}{\partial x^2}} e^{-gx^2} \int_{-\infty}^{\infty} e^{i\left(xP + \frac{P^3}{3}\right)}dP \quad e^{\frac{i\tau}{2} \frac{\partial^2}{\partial y^2}} e^{-gy^2} \int_{-\infty}^{\infty} e^{i\left(yQ + \frac{P^3}{3}\right)}dQ.\label{E3(x,y,0)3D_2}
\end{equation}
We notice that the final expression can be expressed as the product of two components. Each component follows the same structure as Eq.~(\ref{E3(x,tau)D_1}) and can therefore be written in the form of Eq.~(\ref{Airygaussianbeam}). With this result, we finally obtain the expression for the Airy-Gaussian beam, given by
\begin{equation}
      U_3(x,y,\tau)  =  \frac{e^{-\frac{gx^2}{\omega(\tau)}}}{\sqrt{\omega(\tau)}} \int_{-\infty}^{\infty} e^{i\left[ \frac{x}{\omega(\tau)}P - \left( \frac{\tau}{2}- \frac{ig\tau^2}{\omega(\tau)} \right) P^2 + \frac{P^3}{3} \right]}dP \quad \frac{e^{-\frac{gy^2}{\omega(\tau)}}}{\sqrt{\omega(\tau)}} \int_{-\infty}^{\infty} e^{i\left[ \frac{y}{\omega(\tau)}Q - \left( \frac{\tau}{2}- \frac{ig\tau^2}{\omega(\tau)} \right) Q^2 + \frac{Q^3}{3} \right]}dQ,
\end{equation}
which, in terms of the Airy function, can be expressed as
\begin{equation}
\begin{split}
U_3(x,y ,\tau) = & 4\pi^2 \frac{e^{\frac{-g(x^2 + y^2)}{\omega(\tau)}}}{\omega(\tau)} \mathrm{Ai}\left(\frac{x}{\omega(\tau)} - \left( \frac{\tau}{2} - \frac{ig\tau^2}{\omega(\tau)} \right)^2 \right) \mathrm{Ai}\left(\frac{y}{\omega(\tau)} - \left( \frac{\tau}{2} - \frac{ig\tau^2}{\omega(\tau)} \right)^2 \right)  e^{i\left[ \frac{x + y}{\omega(\tau)} \left( \frac{\tau}{2} - \frac{ig\tau^2}{\omega(\tau)} \right) - \frac{4}{3}\left( \frac{\tau}{2}- \frac{ig\tau^2}{\omega(\tau)} \right)^3 \right]}.
\end{split}
\label{AiryGaussianbeam2D}
\end{equation}

\begin{figure}[!ht]
  \centering
  \includegraphics[width=\linewidth]{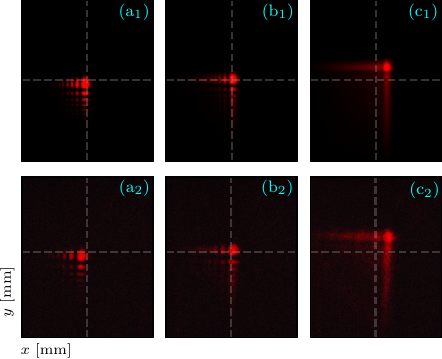}
  \caption{Intensity distributions for the Airy-Gaussian beam at three transverse planes: ($\mathrm{a}_1$) at $z = 0.0 \, \mathrm{m}$, ($\mathrm{b}_1$) at $z = 0.2 \, \mathrm{m}$ and ($\mathrm{c}_1$) at $z = 0.2 \, \mathrm{m}$. The corresponding experimental distributions are shown in ($\mathrm{a}_2$)-($\mathrm{c}_2$), the experimental parameters are $T=1 \times 10^{-4} \, \mathrm{m}$, $g = 0.125 \times 10^{7} \,  \mathrm{m}^{-2}$ and $\lambda = 632.8 \, \mathrm{nm}$, all within a viewing window of $4 \, \mathrm{mm}$. The Gaussian confinement factor provides a smoother energy decay compared to the exponential truncation, enhancing the beam's stability.}
  \label{figplotAiryg2D}
\end{figure}

In Fig. \ref{figplotAiryg2D}, we present the analytical and experimental intensity distribution associated with Eq.~(\ref{AiryGaussianbeam2D}) at different values of $\tau$. For the experimental generation of this third optical field, the experimental arrangement detailed in \label{Propagation(2+1)D-A} was retained. However, the synthetic phase hologram (SPH) encoded on the spatial light modulator (SLM) was suitably updated to project this specific beam. Much like their truncated counterparts, Airy-Gaussian beams confine both amplitude and energy, albeit through a more pronounced modulation profile. Importantly, they still maintain their characteristic parabolic trajectory within the $(x,y)$ plane.\\

Finally, we mention that both the Airy-truncated Eq.~(\ref{Airytruncatedbeam2D}) and Airy-Gaussian Eq.~(\ref{AiryGaussianbeam2D}) beams are physically realizable. The Airy-truncated beam achieves physical realizability through exponential truncation, which confines the energy while preserving the characteristic parabolic self-accelerating trajectory. However, the Airy-Gaussian beam introduces a transverse Gaussian envelope, which limits the amplitude and energy in space, ensuring that the beam can be generated experimentally and that its transverse spreading during propagation can be controlled. Thus, while both beams maintain self-accelerating behavior, the Gaussian modulation provides smoother control and stability during propagation.

\section{Conclusions}
In this work, we have demonstrated that the propagation of complex structured light fields, specifically Airy-type beams, can be elegantly solved using the algebraic formalism of quantum mechanical operators. By exploiting the formal isomorphism between the paraxial wave equation and the Schrödinger equation, we transitioned from the traditionally cumbersome Fresnel diffraction integrals to a streamlined operator-based approach.\\

Our derivation of the $(1+1)D$ and $(2+1)D$ Airy, Airy-truncated, and Airy-Gaussian beams highlights several pedagogical advantages. Firstly, it reinforces the student's understanding of unitary evolution operators, a fundamental concept of quantum mechanics, by applying them to a tangible optical system. Secondly, the use of the Hadamard lemma and the Baker-Campbell-Hausdorff formula provides a clear, step-by-step algebraic path that avoids the complexities of integrating highly oscillatory functions like the Airy function. Finally, the operator method allows students to see the ``transformation'' of the initial field as a sequence of shift and scaling operations, making the physics of self-acceleration and non-diffraction more intuitive. The excellent agreement between our analytical solutions and the experimental intensity profiles obtained with a spatial light modulator (SLM) serves as a powerful validation of this theoretical framework. For the physics student, seeing that abstract non-commuting operators translate directly into the ``curved" light paths observed in the laboratory is a compelling lesson in the power of mathematical physics. We believe that introducing these methods in advanced undergraduate or early graduate optics and quantum mechanics courses can bridge the gap between these two fundamental pillars of physics education, providing students with a more versatile and unified problem-solving toolkit.

\appendix
\section{Alternative way to obtain the propagated Airy beam applying the Hadamard lemma and the Baker-Campbell-Hausdorff formula}\label{app:alternative}

To find the expression of the $(1+1)D$ Airy beam in free space using operator techniques, we consider the following expression as the initial condition,
\begin{equation}
u_1(x,0) = \int_{-\infty}^{\infty} e^{i\left(xP + \frac{P^3}{3}\right)}dP,\label{E1(x,0)D_1App}
\end{equation}
which corresponds to the Airy function given by $2\pi\mathrm{Ai}(x)$. Substituting the initial condition into Eq.~(\ref{propagator1D}), the propagated beam is given by  
\begin{equation}
u_1(x,\tau) = e^{\frac{i\tau}{2}\frac{\partial^2}{\partial x^2}} \int_{-\infty}^{\infty} e^{ixP} e^{\frac{iP^3}{3}}dP.\label{E1(x,tau)D_1App}
\end{equation}
Placing the evolution operator inside the integral and inserting the identity operator $\hat{I} = \exp\left(-\frac{i\tau}{2}\frac{\partial^2}{\partial x^2}\right) \exp\left(\frac{i\tau}{2}\frac{\partial^2}{\partial x^2}\right)$ after the factor $\exp(ixP)$, we have the following, 
\begin{equation}
   u_1(x,\tau) = \int_{-\infty}^{\infty} e^{\frac{i\tau}{2}\frac{\partial^2}{\partial x^2}} e^{ixP} e^{-\frac{i\tau}{2}\frac{\partial^2}{\partial x^2}} e^{\frac{i\tau}{2}\frac{\partial^2}{\partial x^2}} e^{\frac{iP^3}{3}}dP.\label{E1(x,tau)2D_1App}
\end{equation}
It follows from the Taylor expansion of the exponential operator $\exp\left(\frac{i\tau}{2}\frac{\partial^2}{\partial x^2}\right)$ that $\exp\left(\frac{i\tau}{2}\frac{\partial^2}{\partial x^2}\right) \exp\left(\frac{iP^3}{3}\right) = \exp\left(\frac{iP^3}{3}\right)$. The next step consists of applying the Hadamard lemma and the Baker-Campbell-Hausdorff formula \, \cite{Rossmann2002, Hall2013} to the factor $\exp\left(\frac{i\tau}{2}\frac{\partial^2}{\partial x^2}\right) \exp(ixP) \exp\left(-\frac{i\tau}{2}\frac{\partial^2}{\partial x^2}\right)$, resulting in
\begin{equation}
     e^{\frac{i\tau}{2}\frac{\partial^2}{\partial x^2}} e^{ixP} e^{-\frac{i\tau}{2}\frac{\partial^2}{\partial x^2}}  =  e^{i(x + i\tau\partial_x)P} = e^{-\frac{i \tau P^2}{2}} e^{ixP} e^{-\tau P\frac{\partial}{\partial x}}.
\end{equation} 
Using these results, the propagated optical beam can be expressed in the following manner
\begin{equation}
    u_1(x,\tau) = \int_{-\infty}^{\infty} e^{-\frac{i \tau P^2}{2}} e^{ixP}  e^{-\tau P\frac{\partial}{\partial x}} e^{\frac{iP^3}{3}}dP.\label{E1(x,tau)3D_1App}
\end{equation}
Given that $\exp\left(-\tau P\frac{\partial}{\partial x}\right) \exp\left(\frac{iP^3}{3}\right) = \exp\left(\frac{iP^3}{3}\right)$ and, after regrouping the terms in the above expression, we arrive at the final expression of the propagated paraxial beam
\begin{equation}
    u_1(x,\tau) = \int_{-\infty}^{\infty} e^{i \left( xP - \frac{ \tau P^2}{2} + \frac{P^3}{3} \right)}dP.\label{E1(x,tau)4D_1App}
\end{equation}
We remark that the expression for the propagated Airy beam has been obtained in two ways by employing operator techniques. In the first case, we use the fact that $\frac{\partial^2}{\partial x^2} e^{ixP} = -P^2 e^{ixP}$. In the second case, we introduce $\hat{I} = \exp\left(-\frac{i\tau}{2}\frac{\partial^2}{\partial x^2}\right) \exp\left(\frac{i\tau}{2}\frac{\partial^2}{\partial x^2}\right)$, and the Hadamard lemma and the Baker-Campbell-Hausdorff formula are applied to arrive at the same expression, Eq.~(\ref{E1(x,tau)3D_1}).

\section{Proof of the operator factorization formula}\label{app:disentangling}
In order to show Eq. (\ref{DisentanglingIdentity}), we start with the expression
\begin{equation}
    e^{\frac{i\tau}{2}\frac{\partial^2}{\partial x^2}} e^{-g x^2} e^{-\frac{i\tau}{2}\frac{\partial^2}{\partial x^2}} = e^{-g\left(x + i\tau\frac{\partial}{\partial x}\right)^2} = e^{-g\left[x^2 + i\tau \left(x\frac{\partial}{\partial x} + \frac{\partial}{\partial x}x \right) -\tau^2 \frac{\partial^2}{\partial x^2}\right]},
\end{equation}
which must be factorized; that is, we should write it as a product of exponential functions. For this, we note that the operator set $\left\{ x^2, \left(x\frac{\partial}{\partial x} + \frac{\partial}{\partial x}x \right), \frac{\partial^2}{\partial x^2} \right\}$ is commutation-closed \cite{Wei1964}; thus, we have the relation
\begin{equation}
    e^{-g\beta\left[x^2 + i\tau \left(x\frac{\partial}{\partial x} + \frac{\partial}{\partial x}x \right) -\tau^2 \frac{\partial^2}{\partial x^2}\right]}  =  e^{h_1(\beta)x^2} e^{h_2(\beta)\left(x\frac{\partial}{\partial x} + \frac{\partial}{\partial x}x \right)} e^{h_3(\beta)\frac{\partial^2}{\partial x^2}},\label{U(beta)}
\end{equation}
where $\beta$ is a parameter. The functions $h_s(\beta)$ $(s = 1,2,3)$ must be determined to differentiate both sides of Eq. (\ref{U(beta)}) with respect to $\beta$, that is,
\begin{align}
    -g\left[x^2 + i\tau \left(x\frac{\partial}{\partial x} + \frac{\partial}{\partial x}x \right) - \tau^2 \frac{\partial^2}{\partial x^2}\right] \hat{U}(\beta) &= \dot{h}_1(\beta) x^2 e^{h_1(\beta)x^2} e^{h_2(\beta)\left(x\frac{\partial}{\partial x} + \frac{\partial}{\partial x}x \right)} e^{h_3(\beta)\frac{\partial^2}{\partial x^2}} \nonumber \\
    &\quad + \dot{h}_2(\beta) e^{h_1(\beta)x^2} \left(x\frac{\partial}{\partial x} + \frac{\partial}{\partial x}x \right) e^{h_2(\beta)\left(x\frac{\partial}{\partial x} + \frac{\partial}{\partial x}x \right)} e^{h_3(\beta)\frac{\partial^2}{\partial x^2}} \nonumber \\
    &\quad + \dot{h}_3(\beta) e^{h_1(\beta)x^2} e^{h_2(\beta)\left(x\frac{\partial}{\partial x} + \frac{\partial}{\partial x}x \right)} \frac{\partial^2}{\partial x^2} e^{h_3(\beta)\frac{\partial^2}{\partial x^2}}, \nonumber \\
    &= \Bigg[ \dot{h}_1(\beta) x^2 + \dot{h}_2(\beta) e^{h_1(\beta)x^2} \left(x\frac{\partial}{\partial x} + \frac{\partial}{\partial x}x \right) e^{-h_1(\beta)x^2} \nonumber \\
    &\quad + \dot{h}_3(\beta) e^{h_1(\beta)x^2} e^{h_2(\beta)\left(x\frac{\partial}{\partial x} + \frac{\partial}{\partial x}x \right)} \frac{\partial^2}{\partial x^2} e^{-h_2(\beta)\left(x\frac{\partial}{\partial x} + \frac{\partial}{\partial x}x \right)} e^{-h_1(\beta)x^2} \Bigg] \hat{U}(\beta),\label{dU(beta)1}
\end{align}
where $\hat{U}(\beta)$ is given on either side by the equality in Eq. (\ref{U(beta)}). In order to continue, we use the Hadamard lemma and the Baker-Campbell-Hausdorff formula; thus,
\begin{eqnarray}
    e^{h_1(\beta)x^2}  \left(x\frac{\partial}{\partial x} + \frac{\partial}{\partial x}x \right) e^{-h_1(\beta)x^2} & = & \left(x\frac{\partial}{\partial x} + \frac{\partial}{\partial x}x \right) - 4h_1(\beta) x^2,\label{identity1}\\
    e^{h_1(\beta)x^2} e^{h_2(\beta)\left(x\frac{\partial}{\partial x} + \frac{\partial}{\partial x}x \right)} \frac{\partial^2}{\partial x^2} e^{-h_2(\beta)\left(x\frac{\partial}{\partial x} + \frac{\partial}{\partial x}x \right)} e^{-h_1(\beta)x^2} & = & e^{-4h_2(\beta)} \left[ \frac{\partial^2}{\partial x^2} - 2h_1(\beta) \left(x\frac{\partial}{\partial x} + \frac{\partial}{\partial x}x \right) +4h_1^2(\beta)x^2\right].\label{identity2} 
\end{eqnarray}
Substituting Eqs. (\ref{identity1}) and (\ref{identity2}) into Eq. (\ref{dU(beta)1}), we find the following: 
\begin{align}
    -g\beta\left[x^2 + i\tau \left(x\frac{\partial}{\partial x} + \frac{\partial}{\partial x}x \right) -\tau^2 \frac{\partial^2}{\partial x^2}\right] \hat{U}(\beta) &= \Bigg\{ \dot{h}_1(\beta) x^2 + \dot{h}_2(\beta) \left[ \left(x\frac{\partial}{\partial x} + \frac{\partial}{\partial x}x \right) - 4h_1(\beta) x^2 \right] \nonumber \\
    &\quad + \dot{h}_3(\beta) e^{-4h_2(\beta)} \left[ \frac{\partial^2}{\partial x^2} - 2h_1(\beta) \left(x\frac{\partial}{\partial x} + \frac{\partial}{\partial x}x \right) + 4h_1^2(\beta)x^2 \right] \Bigg\} \hat{U}(\beta). \label{dU2(beta)2}
\end{align}
For Eq.\,(\ref{U(beta)}) to hold, Eq.\,(\ref{dU2(beta)2}) must be satisfied term by term. This requirement yields the following system of coupled differential equations for the functions $h_s(\beta)$:
\begin{eqnarray}
    \dot{h}_1(\beta) - 4h_1(\beta) \dot{h}_2(\beta) + 4h_1^2(\beta) \dot{h}_3(\beta) e^{-4h_2(\beta)} & = & - g,\\
    \dot{h}_2(\beta) - 2h_1(\beta) \dot{h}_3(\beta) e^{-4h_2(\beta)} & = & -ig\tau,\\
    \dot{h}_3(\beta) e^{-4h_2(\beta)} & = & g\tau^2.
\end{eqnarray}
We are interested in the solutions for $h_s(\beta)$ when $\beta = 1$. In this case, we have that
\begin{equation}\label{functionshs}
    h_1(\beta =1) = -\frac{g}{\omega(\tau)} ,\qquad  h_2(\beta = 1) = \frac{i\pi}{4} - \frac{1}{2}\ln[i\omega(\tau)] , \qquad h_3(\beta = 1) = \frac{g\tau^2}{\omega(\tau)},
\end{equation}
with $\omega(\tau) = 1+2ig\tau$.

\end{document}